\documentclass[prd,showpacs]{revtex4}
\begin{document}
\title{Magnetic field corrections to the repulsive Casimir effect at finite temperature}
\author{Andrea Erdas}
\email{aerdas@loyola.edu}
\affiliation{Department of Physics, Loyola University Maryland, 4501 North Charles Street,
Baltimore, Maryland 21210, USA}
\begin {abstract} 
I investigate the finite temperature Casimir effect for a charged and massless  scalar field satisfying mixed (Dirichlet-Neumann) boundary conditions on a pair of 
plane parallel plates of infinite size. The effect of a uniform magnetic field, perpendicular to the plates, on the Helmholtz free energy and Casimir pressure is studied. 
The $\zeta$-function regularization technique is used to obtain finite results. Simple analytic expressions are obtained for the zeta function
and the free energy, in the limits of small plate distance, high temperature and strong magnetic field. The Casimir pressure
is obtained in each of the three limits and the situation of a magnetic field present between and outside the plates, 
as well as that of a magnetic field present only between the plates is examined. It is discovered that, in the small plate distance and high
temperature limits, the repulsive pressure is less
when the magnetic field is present between the plates but not outside, than it is when the magnetic field is present between and outside the plates.
\end {abstract}
\maketitle
\section{Introduction}
\label{1}
The attractive force between two conducting and
electrically neutral  parallel plates in a vacuum was first predicted theoretically by Casimir  \cite{Casimir:1948dh}. 
The first experimental evidence of the attractive Casimir force dates back more than fifty years \cite{Sparnaay:1958wg},
and many followed it. A comprehensive list 
of these experiments is available in the review article and book by Bordag et al. \cite{Bordag:2001qi,Bordag:2009zz}.

Years later, the repulsive spherical Casimir effect was discovered by Boyer, who showed 
that a conducting and electrically neutral  spherical shell in a vacuum is subject to an outward pressure caused
by the quantum fluctuations of the electromagnetic field \cite{Boyer:1968uf}. 
The repulsive Casimir effect for two parallel plates was also discovered by Boyer, within the framework of random 
electrodynamics, when he showed that two electrically neutral
parallel plates in a vacuum, one perfectly
conducting ($\epsilon \rightarrow 0$) and the other infinitely permeable ($\mu \rightarrow \infty$), are subject to a repulsive 
force \cite{Boyer:1974}. This problem has been revisited more recently by several authors employing
modern regularization methods, such as the zeta function technique \cite{Elizalde:1988rh,Elizalde:2007du}. 
This technique has been used 
to calculate the finite temperature corrections to the repulsive 
Casimir effect within the framework of finite temperature field theory, in the case of a massless scalar field 
that mimics the electromagnetic field \cite{Santos:1998vb,CougoPinto:1998xn,Santos:1999yj}. These authors impose 
Dirichlet boundary conditions for the scalar field on one plate and Neumann boundary conditions on the other plate,
simulating a perfectly conducting and an infinitely permeable plate respectively, and assume this system to be in 
thermal equilibrium with a heat reservoir. The massless scalar field with mixed boundary conditions is 
equivalent to the electromagnetic system studied by Boyer in \cite{Boyer:1968uf}, and yields the same results if one 
accounts correctly for the two electromagnetic polarization states.

In this paper I will study the effect of a uniform magnetic field $\vec B$ on the repulsive Casimir effect, by investigating a system very similar
to the one studied in Refs. \cite{Santos:1998vb,CougoPinto:1998xn,Santos:1999yj}; a massless, but charged, scalar field 
satisfying mixed boundary conditions on two plane parallel plates of infinite size and distance $a$, 
and in thermal equilibrium with a heat reservoir at temperature $T$. While several
authors have investigated thermal and magnetic corrections  to the attractive Casimir effect, see for example
\cite{CougoPinto:1998td,CougoPinto:1998jg,Erdas:2013jga}, a calculation of thermal and magnetic corrections to the 
repulsive Casimir effect associated with Boyer's setup has not been done. Boyer's pair of plates is the simplest
system where we can observe the repulsive Casimir force at work. The spherical Casimir effect, also repulsive, requires
very complex calculations for obtaining the Casimir force alone \cite{Bender:1994zr,Cognola:1999zx}, without including thermal or magnetic corrections. While 
thermal effects have been investigated within the context of the spherical 
geometry \cite{Balian:1977qr,Teo:2013bza},
magnetic field effects have never been investigated. A solid 
understanding of magnetic and thermal effects for the Boyer setup will help understand the roles played by 
magnetic fields and temperature effects in other systems where the Casimir force is repulsive but the 
different geometry complicates the calculations significantly, such as the spherical system.

In Sec \ref{2} I calculate  the zeta function for this system, exact to all orders in $eB$, $T$ and $a$, where 
$e$ is the scalar field charge. In Sec. \ref{3} I examine the small plate distance limit ($a^{-1}\gg T,  \sqrt{eB}$)
and obtain a simple analytic expression for the zeta function and the free energy. In Sec. \ref{4} I obtain simple
analytic expressions  for the zeta function and 
free energy in the high temperature limit ($T \gg a^{-1},  \sqrt{eB}$).
In Sec. \ref{5} I examine the strong magnetic field limit ($ \sqrt{eB} \gg a^{-1}, T$)  and derive 
simple analytic expressions for the zeta function and the free energy. In Sec. \ref{6} I obtain simple
analytic expressions for the Casimir pressure 
in the small plate distance limit, the high temperature limit, and the strong magnetic field limit. I also examine the two 
different scenarios where a) the magnetic field is present 
between the plates as well as outside the plates, and b)
where the magnetic field is only present between the plates, and obtain the Casimir pressure, in each of the three limits,
 for both scenarios.
The conclusions with a discussion of the results of this work are presented in Sec. \ref{7}.
\section{Evaluation of the zeta function}
\label{2}

For a system in thermal equilibrium with a heat reservoir, the imaginary time formalism of finite temperature field theory is convenient, and it allows
only field configurations satisfying the following
\begin{equation}
\phi(x,y,z,\tau)=\phi(x,y,z,\tau+\beta),
\label{temperature}
\end{equation}
for any $\tau$, where $\beta=1/T$ is the periodic length in the Euclidean time axis. In addition to the finite temperature boundary condition
(\ref{temperature}), I impose Dirichlet boundary conditions on one plate and Neumann boundary conditions on the other.
The two plates are perpendicular to the $z$-axis
and located at $z=0$ and $z=a$. Dirichlet boundary conditions on the plate at $z=0$, constrain the scalar field
to vanish at that plate,
\begin{equation}
\phi(x,y,0,\tau)=0,
\label{Dirichlet}
\end{equation}
Neumann boundary conditions on the plate at $z=a$, constrain the derivative of the scalar field
to vanish at that plate,
\begin{equation}
{\partial\phi\over\partial z}(x,y,a,\tau)=0.
\label{Neumann}
\end{equation}

In the slab region there is also a uniform magnetic field pointing in the $z$ direction, ${\vec B}=(0,0,B)$. The scalar field
has charge $e$ and interacts with the magnetic field.

The scalar field Helmholtz free energy is
$$F=\beta^{-1}\log \,\det \left(D_{\rm E}|{\cal F}_a\right),$$
where the symbol ${\cal F}_a$ indicates the set of eigenfunctions of the operator $D_{\rm E}$ which satisfy boundary conditions 
(\ref{temperature}),  (\ref{Dirichlet})  and (\ref{Neumann}), and $D_{\rm E}$ is defined as
$$D_{\rm E} = \partial^2_\tau+\partial^2_z-({\vec p} -e{\vec A})^2_\perp,
$$
where $\vec A$ is the electromagnetic vector potential, the subscript $E$ indicates Euclidean time, and I use
the notation ${\vec p}_\perp=(p_x,p_y,0)$.

The zeta function technique allows me to use the eigenvalues of $D_{\rm E}$ to evaluate the free energy.
The mixed boundary conditions (\ref{Dirichlet}) and (\ref{Neumann})  are satisfied only
if the allowed values for the momentum in the $z$ direction are
\begin{equation}
p_z={\pi\over a}\left(n+\textstyle{\frac{1}{2}}\right),
\label{pz}
\end{equation}
where $n \in \{0, 1, 2, 3,...\}$. The eigenvalues of  
$({\vec p} -e{\vec A})^2_\perp$ are the Landau levels
\begin{equation}
2eB\left(l+\textstyle{\frac{1}{2}}\right),
\label{eigenvalues2}
\end{equation}
with $l \in \{0, 1, 2, 3,...\}$. Using the eigenvalues (\ref{pz}) and (\ref{eigenvalues2}), I construct
the zeta function $\zeta\left(s\right)$ of $D_{\rm E}$
$$\zeta(s)= 
{L^2} \sum_{n=0}^\infty \sum_{m=-\infty}^\infty\left(
{eB \over 2\pi}\right) \sum_{l=0}^\infty\mu^{2s}\left[
{\pi^2\over a^2}\left(n+\textstyle{\frac{1}{2}}\right)^2+{4\pi^2\over\beta^2}m^2+eB\left(2l+1\right)
\right]^{-s},
$$
where the factor $eB/ 2\pi$ takes into account the degeneracy per unit 
area of the Landau levels, $L^2$ is the area of the plates, and the parameter $\mu$ with
dimension of mass keeps $\zeta(s)$ dimensionless for all values of $s$. 

Once I will obtain a suitable closed form for the operator $\zeta(s)$,
I will find the free energy by taking a simple derivative of $\zeta(s)$
\begin{equation}
F=-\beta^{-1}\zeta'(0).
\label{Fandzeta}
\end{equation}
With the help of the following identities
$$z^{-s}={1\over \Gamma(s)}\int_0^\infty dt\, t^{s-1}e^{-zt},$$
\begin{equation}
\sum_{l=0}^\infty e^{-(2l+1)z}={1\over 2 \sinh z},
\label{sinh}
\end{equation}
where $\Gamma (s)$ is the Euler gamma function, I rewrite $\zeta(s)$ as
\begin{equation}
\zeta(s)= {L^2\mu^{2s} \over 4\pi\Gamma(s)}
\int_0^\infty dt \, t^{s-2} {eBt \over \sinh eBt}\left(\sum_{n=0}^\infty e^{-{\pi^2\over a^2}\left(n+{\frac{1}{2}}\right)^2 t}\right)
\left(\sum_{m=-\infty}^\infty e^{-{4\pi^2\over \beta^2}m^2 t}\right).
\label{zeta2}
\end{equation}
It is not possible to evaluate (\ref{zeta2}) in closed form for arbitrary values of $B$, $a$ and $\beta$, but it 
is possible to find simple expressions for $\zeta(s)$  when one or some of $B$, $a$ and $T$ are small or large. 
From these simple expressions of the zeta function,  the free energy and the Casimir pressure will be obtained easily.

\section{Small plate distance}
\label{3}

I will first evaluate $\zeta(s)$ in the small plate distance limit ($a^{-1}\gg T, \sqrt{eB}$). In order to do that, I apply
Poisson's resummation formula  \cite{Dittrich:1979ux} to 
the $m$ sum in (\ref{zeta2}) and find
\begin{equation}
\zeta(s)=\tilde{\zeta}_{B,a,T}(s)+\tilde{\zeta}_{B,a}(s) ,
\label{z}
\end{equation}
where
\begin{equation}
\tilde{\zeta}_{B,a,T}(s)=
{L^2 \mu^{2s}\beta\over 4\pi^{3/2}\Gamma(s)}\sum_{n=0}^\infty \sum_{m=1}^\infty 
\int_0^\infty dt \, t^{s-5/2} {eBt \over \sinh eBt} e^{-m^2\beta^2/ 4t}
e^{-{\pi^2\over a^2}\left(n+{\frac{1}{2}}\right)^2 t},
\label{zBaT}
\end{equation}
\begin{equation}
\tilde{\zeta}_{B,a}(s)=
{L^2\mu^{2s} \beta\over 8\pi^{3/2}\Gamma(s)}\sum_{n=0}^\infty 
\int_0^\infty dt \, t^{s-5/2} {eBt \over \sinh eBt} e^{-{\pi^2\over a^2}\left(n+{\frac{1}{2}}\right)^2 t}.
\label{zBa}
\end{equation}
The zeta function of Eq. (\ref{z}) is equivalent to (\ref{zeta2}), but better suited for a small plate distance expansion.

After changing 
the integration variable from $t$ to $t m a\beta /\pi \left(2n+1\right)$ in  (\ref{zBaT}), I obtain
$$\tilde{\zeta}_{B,a,T}(s)=
{L^2\mu^{2s}\beta \over 4\pi^{3/2}\Gamma(s)}\sum_{n=0}^\infty \sum_{m=1}^\infty \left[{ma\beta\over \pi  \left(2n+1\right)}\right]^{s-1/2}
\int_0^\infty dt \, t^{s-5/2} {eBt \over \sinh \left[{eBtma\beta\over \pi (2n+1)}\right]} e^{-\pi \left(n+{\frac{1}{2}}\right)m\beta(t+1/t)/2a}.
$$
When $aT \ll1$, only the term with $n=0$ and $m=1$ contributes significantly to the double sum so, 
using the saddle point method, I evaluate the integral and obtain
\begin{equation}
\tilde{\zeta}_{B,a,T}(s)=
{L^2 eB \over 2\pi \Gamma(s)} \left({a\beta\mu^2\over \pi}\right)^{s}
{ e^{-\pi \beta/2a}\over \sinh ({eBa\beta\over \pi})}.
\label{zBaT3}
\end{equation}
Next I evaluate (\ref{zBa}) for $a\sqrt{eB}\ll 1$. In this case,  I can set 
\begin{equation}
{eBt \over\sinh eBt}\approx 1-{1\over 6} (eBt)^2
\label{smallB}
\end{equation}
and, after substituting (\ref{smallB}) into (\ref{zBa}),  I integrate to find
\begin{equation}
\tilde{\zeta}_{B,a}(s) =
{\pi^{3/2}L^2 \beta\over 8a^3\Gamma(s)}\left({a\mu\over \pi}\right)^{2s}
\left[\Gamma(s-{\textstyle{3\over 2}})\zeta_H(2s-3,{\textstyle\frac{1}{2}})-
{e^2B^2a^4\over 6\pi^{4}}\Gamma(s+{\textstyle{1\over 2}})\zeta_H(2s+1,{\textstyle\frac{1}{2}})\right],
\label{ztBT2}
\end{equation}
where
$$\zeta_H(s,a)=\sum_{n=0}^\infty (n+a)^{-s}
$$
is the Hurwitz zeta function. 

To
calculate the free energy, it is sufficient to know $\zeta(s)$ for $s\rightarrow 0$. For small $s$ I find
\begin{equation}
{z^{s}\over \Gamma(s)}=s +{\cal O}(s^2),
\label{lim0}
\end{equation}
\begin{equation}
z^{2s}\zeta_H(2s-3,{\textstyle\frac{1}{2}}){\Gamma(s-{\scriptstyle{3\over 2}})\over \Gamma(s)}=-{7\sqrt{\pi}\over 720}s +{\cal O}(s^2),
\label{lim1}
\end{equation}
and
\begin{equation}
z^{2s}\zeta_H(2s+1,{\textstyle\frac{1}{2}}){\Gamma(s+{\scriptstyle{1\over 2}})\over \Gamma(s)}={\sqrt{\pi}\over 2} 
+\sqrt{\pi}\left[\gamma_E+\ln ({2z})\right]s+{\cal O}(s^2),
\label{lim2}
\end{equation}
where $\gamma_E = 0.5772$ is the Euler Mascheroni constant. 
Substituting (\ref{lim0}), and (\ref{lim1}) - (\ref{lim2}) into (\ref{zBaT3}) and (\ref{ztBT2}) respectively, I obtain
$$\zeta(s)=-
L^2\beta\left[{7\pi^{2}\over 5,760a^3}+
{e^2B^2a\over 48\pi^{2}}\left({1\over 2s}+\gamma_E + \ln{2a\mu\over \pi}\right)-{ eB \over 2\pi \beta} 
{ e^{-\pi \beta/2a}\over \sinh ({eBa\beta\over \pi})}\right]s,
$$
valid in the limit of small plate distance and small $s$. The free energy in the small plate distance limit is obtained immediately using (\ref{Fandzeta})
\begin{equation}
F=
L^2\left[{7\pi^{2}\over 5,760a^3}+
{e^2B^2a\over 48\pi^{2}}\left(\gamma_E + \ln{2a\mu\over \pi}\right)-{ eB \over 2\pi \beta} 
{ e^{-\pi \beta/2a}\over \sinh ({eBa\beta\over \pi})}\right].
\label{F0}
\end{equation}
\section{High temperature}
\label{4}

Next I evaluate $\zeta(s)$ in the high temperature limit ($T \gg a^{-1}, \sqrt{eB}$), and apply
Poisson's resummation formula to the $n$ sum in (\ref{zeta2}) to find
\begin{equation}
\zeta(s)={\zeta}_{B,a,T}(s)+{\zeta}_{B,a}(s)+{\zeta}_{B,T}(s) ,
\label{z0}
\end{equation}
where
\begin{equation}
{\zeta}_{B,a,T}(s)=
{L^2 \mu^{2s} a\over 2\pi^{3/2}\Gamma(s)}\sum_{n=1}^\infty \sum_{m=1}^\infty (-1)^n
\int_0^\infty dt \, t^{s-5/2} {eBt \over \sinh eBt}
e^{-{4\pi^2\over \beta^2}m^2 t} e^{-n^2a^2/ t},
\label{zBaT0}
\end{equation}
$$
{\zeta}_{B,T}(s)=
{L^2 \mu^{2s} a\over 8\pi^{3/2}\Gamma(s)} \sum_{m=-\infty}^\infty 
\int_0^\infty dt \, t^{s-5/2} {eBt \over \sinh eBt}
e^{-{4\pi^2\over \beta^2}m^2 t},
$$
$$
{\zeta}_{B,a}(s)=
{L^2 \mu^{2s} a\over 4\pi^{3/2}\Gamma(s)}\sum_{n=1}^\infty (-1)^n
\int_0^\infty dt \, t^{s-5/2} {eBt \over \sinh eBt}
e^{-n^2a^2/ t}.
$$
Notice that $\zeta(s)$ from Eq. (\ref{z0}), while equivalent to (\ref{zeta2}) and (\ref{z}), it is better suited for a high 
temperature expansion.

I evaluate (\ref{zBaT0}) by changing the variable of integration $(t\rightarrow ta\beta n/2\pi m)$, and obtain
$${\zeta}_{B,a,T}(s)=
{L^2 \mu^{2s} a\over 2\pi^{3/2}\Gamma(s)}\sum_{n=1}^\infty \sum_{m=1}^\infty (-1)^n  \left({a\beta n\over 2\pi  m}\right)^{s-1/2}
\int_0^\infty dt \, t^{s-5/2} {eBt \over \sinh \left({eBta\beta n\over 2\pi m}\right)} e^{-2\pi nma(t+1/t)/\beta}.
$$
When $aT\gg 1$, all terms in the double sum are negligible when compared to the $m=n=1$ term and, using the 
saddle point method, I find
$$
{\zeta}_{B,a,T}(s)=
-{L^2 eB \over 2\pi \Gamma(s)} \left({a\beta\mu^2\over 2\pi}\right)^{s}
{ e^{-4\pi a/\beta}\over \sinh ({eBa\beta\over 2\pi})}.
$$

For $T\gg \sqrt{eB}$, I use (\ref{smallB}), and find
$${\zeta}_{B,T}(s)=
{L^2 \mu^{2s} a\over 8\pi^{3/2}\Gamma(s)} 
\int_0^\infty dt \, t^{s-5/2} \left[{eBt \over \sinh eBt}+2\sum_{m=1}^\infty 
\left(1-{e^2B^2t^2\over 6}\right)e^{-{4\pi^2\over \beta^2}m^2 t}\right],
$$
and, after integration
\begin{eqnarray}
{\zeta}_{B,T}(s)=
{L^2 a\over \Gamma(s)}&&
\!\!\!\!\!\! \left[\left({eB\over 2\pi}\right)^{3/2}\left({\mu^2\over 2eB}\right)^s\Gamma(s-{\textstyle\frac{1}{2}})
 \zeta_H(s-{\textstyle\frac{1}{2}},{\textstyle\frac{1}{2}})+{2\pi^{3/2}\over \beta^3}
 \left({\mu \beta\over 2\pi}\right)^{2s}\Gamma(s-{\textstyle\frac{3}{2}})
 \zeta_R(2s-3)\right.
\nonumber \\
&&
\!\!\!\left.- {e^2B^2\beta\over48\pi^{5/2}}\left({\mu \beta\over 2\pi}\right)^{2s}\Gamma(s+{\textstyle\frac{1}{2}})
 \zeta_R(2s+1)
 \right],
\nonumber 
\end{eqnarray}
where $\zeta_R(s)$ is the Riemann zeta function.

Last I calculate ${\zeta}_{B,a}(s)$, which contains only two of the parameters, $B$ and $a$. In the high temperature limit, 
we need to explore the cases of $\sqrt{eB}\gg a^{-1}$ and of $a^{-1}\gg\sqrt{eB}$ separately. When $\sqrt{eB}\gg a^{-1}$, I can take
$$({\sinh eBt})^{-1}\approx 2e^{-eBt},
$$
and write
$${\zeta}_{B,a}(s)=
{L^2 \mu^{2s} a\over 2\pi^{3/2}\Gamma(s)}\sum_{n=1}^\infty (-1)^n
\int_0^\infty dt \, t^{s-5/2} {eBt }e^{-eBt}
e^{-n^2a^2/ t},
$$
which, after a change of the integration variable, becomes
$${\zeta}_{B,a}(s)=
{L^2 \mu^{2s} aeB\over 2\pi^{3/2}\Gamma(s)}\sum_{n=1}^\infty (-1)^n
\left({an\over\sqrt{eB}}\right)^{s-1/2}\int_0^\infty dt \, t^{s-3/2} 
e^{-na\sqrt{eB}(t+1/ t)}.
$$
When $a\sqrt{eB}\gg 1$, only the term with $n=1$ survives in the sum, and the integration is done quickly using the 
saddle point method to obtain
$$
{\zeta}_{B,a}(s)=
-{L^2  eB\over 2\pi\Gamma(s)}
\left({\mu^2a\over\sqrt{eB}}\right)^{s}
e^{-2a\sqrt{eB}}.
$$
In the case of high temperature and $a^{-1}\gg\sqrt{eB}$, I use  (\ref{smallB}), integrate and obtain
$${\zeta}_{B,a}(s)=
{L^2 (\mu a)^{2s} \over 4\pi^{3/2}a^2\Gamma(s)}\left[(2^{2s-2}-1)\zeta_R(3-2s)\Gamma({\textstyle\frac{3}{2}}-s)-
{e^2B^2a^4\over 6}(2^{2s+2}-1)\zeta_R(-1-2s)\Gamma(-{\textstyle\frac{1}{2}}-s)
\right].
$$

The small $s$ expansion of $\zeta(s)$, in the high temperature limit, is
$$
{\zeta}(s)=
{L^2 a \over 2\pi}\left[
{2\pi^3\over 45\beta^3}+(eB)^{3/2}(\sqrt{2}-1)\zeta_R(-{\textstyle\frac{1}{2}})
-{e^2B^2\beta\over 6}\left({1\over 2s}+\gamma_E+\ln{\mu\beta\over4 \pi}\right)
-{eB\over a}e^{-2a\sqrt{eB}}
-{ eBe^{-4\pi a/\beta}\over a\sinh ({eBa\beta\over 2\pi})}
\right]s,
$$
for $\sqrt{eB}\gg a^{-1}$, and
$$
{\zeta}(s)=
{L^2 a \over 2\pi}\left[
{2\pi^3\over 45\beta^3}+(eB)^{3/2}(\sqrt{2}-1)\zeta_R(-{\textstyle\frac{1}{2}})
-{e^2B^2\beta\over 6}\left({1\over 2s}+\gamma_E+\ln{\mu\beta\over4 \pi}\right)
-{3\zeta_R(3)\over 16 a^3}- {e^2B^2 a\over 24}
-{ eBe^{-4\pi a/\beta}\over a\sinh ({eBa\beta\over 2\pi})}
\right]s,
$$
for $a^{-1}\gg \sqrt{eB}$. The free energy in the high temperature limit is obtained using (\ref{Fandzeta}),
\begin{equation}
F=
-{L^2 a \over 2\pi}\left[
{2\pi^3\over 45\beta^4}+{(eB)^{3/2}\over \beta}(\sqrt{2}-1)\zeta_R(-{\textstyle\frac{1}{2}})
-{e^2B^2\over 6}\left(\gamma_E+\ln{\mu\beta\over4 \pi}\right)
-{eB\over \beta a}e^{-2a\sqrt{eB}}
-{ eBe^{-4\pi a/\beta}\over \beta a\sinh ({eBa\beta\over 2\pi})}
\right]
\label{zBa020a}
\end{equation}
for $\sqrt{eB}\gg a^{-1}$, and
\begin{equation}
F=
-{L^2 a \over 2\pi}\left[
{2\pi^3\over 45\beta^4}+{(eB)^{3/2}\over \beta}(\sqrt{2}-1)\zeta_R(-{\textstyle\frac{1}{2}})
-{e^2B^2\over 6}\left({a\over 4\beta}+\gamma_E+\ln{\mu\beta\over4 \pi}\right)
-{3\zeta_R(3)\over 16 \beta a^3}
-\left({2\pi\over a^2 \beta^2}-{e^2B^2\over 12\pi}\right){ e^{-4\pi a/\beta}}
\right]
\label{zBa021a}
\end{equation}
for $a^{-1}\gg \sqrt{eB}$, where I used a power series expansion of the hyperbolic sine because, for high temperature 
and $a^{-1}\gg \sqrt{eB}$, the quantity $eBa\beta \ll1$.
\section{Strong magnetic field}
\label{5}
In the strong magnetic field limit ($\sqrt{eB} \gg a^{-1}, T$), a form of the zeta function that can be easily 
expanded is obtained by Poisson-resumming over both $m$ and $n$ in (\ref{zeta2})
$$\zeta(s)= {L^2a\beta \mu^{2s}\over 16\pi^2\Gamma(s)}\sum_{m,n=-\infty}^\infty(-1)^n
\int_0^\infty dt \, t^{s-3} {eBt \over \sinh eBt}
e^{- {\beta^2 m^2 \over 4t}}e^{- {a^2 n^2 \over t}},
$$
and, using (\ref{sinh}), is rewritten as
$$
\zeta(s)= {L^2a\beta  \mu^{2s}\over 8\pi^2\Gamma(s)}\sum_{m,n=-\infty}^\infty \sum_{l=0}^\infty(-1)^n eB
\int_0^\infty dt \, t^{s-2}  e^{-(2l+1)eBt}
e^{- {\beta^2 m^2 \over 4t}}e^{- {a^2 n^2 \over t}}.
$$
After changing the integration variable, I obtain
$$
\zeta(s)=\zeta_W(s)+ {\tilde\zeta}(s),
$$
with
\begin{equation}
\zeta_W(s)= {L^2a\beta  \mu^{2s}\over 8\pi^2\Gamma(s)}\sum_{l=0}^\infty eB
\int_0^\infty dt \, t^{s-2}  e^{-(2l+1)eBt},
\label{zBB3}
\end{equation}
and
\begin{equation}
{\tilde\zeta}(s)= {L^2a\beta  \mu^{2s}\over 8\pi^2\Gamma(s)}{\sum_{m,n}}'\sum_{l=0}^\infty(-1)^n eB
\left({\beta^2m^2/4+a^2n^2\over (2l+1)eB}\right)^{(s-1)/2}
\int_0^\infty dt \, t^{s-2}  e^{-(t+1/t)\sqrt{(2l+1)eB({\beta^2 m^2 \over 4}+a^2n^2)}},
\label{zBB}
\end{equation}
where the primed summation over $n$ and $m$ does not include the term with $n=m=0$. Notice that $\zeta_W(s)$, as defined in Eq.
(\ref{zBB3}), yields the zeta function of the one-loop Weisskopf effective Lagrangian for scalar QED  \cite{Weisskopf:1996bu}, 
in the limit of a massless scalar.
The integral present in
(\ref{zBB3}) is evaluated easily to obtain
\begin{equation}
\zeta_W(s)= {L^2a\beta(eB)^2  \over 4\pi^2\Gamma(s)}\left({\mu^{2}\over 2eB}\right)^s
\zeta_H(s-1,{\textstyle\frac{1}{2}})\Gamma(s-1).
\label{zBB4}
\end{equation}
Once I compare Eq. (\ref{zBB4}) to the well known result for the Weisskopf Lagrangian \cite{Weisskopf:1996bu}, I realize that I 
must take the arbitrary parameter $\mu=m_\phi$, where $m_\phi$ is the small mass of the scalar field, negligible when compared to
the three relevant physical quantities $a$, $T$ and $\sqrt{eB}$. 

The integral in (\ref{zBB}) is done using the saddle point method, to obtain
\begin{equation}
{\tilde\zeta}(s)= {L^2a\beta  \mu^{2s}\over 8\pi^{3/2}\Gamma(s)}{\sum_{m,n}}'\sum_{l=0}^\infty(-1)^n eB
\left({\beta^2m^2/4+a^2n^2\over (2l+1)eB}\right)^{s/2}{[(2l+1)eB]^{1/4}\over[\beta^2m^2/4+a^2n^2]^{3/4}}
e^{-\sqrt{(2l+1)eB({\beta^2 m^2 }+4a^2n^2)}}.
\label{zBB4a}
\end{equation}
Only terms with $l=0$, $m=\pm1$, $n=0$; or   $l=1$, $m=\pm1$, $n=0$; or $l=0$, $m=0$, $n=\pm 1$ contribute significantly to the
sum in Eq. (\ref{zBB4a}), and I obtain
$$
{\tilde\zeta}(s)= {L^2a\beta (eB)^{5/4}  \over 4\pi^{3/2}\Gamma(s)} \left[
\left({\mu^2\beta\over 2\sqrt{eB}}\right)^s{\left(2\over\beta\right)^{3/2}}
e^{-\beta\sqrt{eB}}+
\left({\mu^2\beta\over 2\sqrt{3eB}}\right)^s{\left(2\sqrt{3}\over\beta\right)^{3/2}}
e^{-\beta\sqrt{3eB}}-
\left({\mu^2a\over \sqrt{eB}}\right)^s{\left(1\over a\right)^{3/2}}
e^{-2a\sqrt{eB}}\right].
$$
In the strong magnetic field limit, the small $s$ expansion of $\zeta(s)$ is
$$
\zeta(s)= {L^2a\beta (eB)^{5/4}\over 4\pi^{3/2}}\left[{(eB)^{3/4}\over 24\sqrt{\pi}}\left(\ln{eB\over 3 \mu^2}-{1\over 2}\right)+
{\left(2\over\beta\right)^{3/2}}
e^{-\beta\sqrt{eB}}+
{\left(2\sqrt{3}\over\beta\right)^{3/2}}
e^{-\beta\sqrt{3eB}}-
{\left(1\over a\right)^{3/2}}
e^{-2a\sqrt{eB}}\right]s,
$$
and therefore, using (\ref{Fandzeta}), I obtain the free energy in the strong magnetic field limit
\begin{equation}
F= -{L^2a (eB)^{5/4}\over 4\pi^{3/2}}\left[{(eB)^{3/4}\over 24\sqrt{\pi}}\left(\ln{eB\over 3m_\phi^2}-{1\over 2}\right)+
{\left(2\over\beta\right)^{3/2}}
e^{-\beta\sqrt{eB}}+
{\left(2\sqrt{3}\over\beta\right)^{3/2}}
e^{-\beta\sqrt{3eB}}
-{\left(1\over a\right)^{3/2}}
e^{-2a\sqrt{eB}}\right].
\label{zBB7}
\end{equation}
Notice that the arbitrary parameter $\mu$ has been replaced by $m_\phi$, and this replacement should also occur in Eqs.
(\ref{F0}), (\ref{zBa020a}),  and (\ref{zBa021a}), the other three expressions I obtained for the free energy.

\section{Casimir pressure}
\label{6}
The Casimir pressure on the plates is given by
\begin{equation}
P=-{1\over L^2}{\partial F\over \partial a}.
\label{P1}
\end{equation}
When calculating the pressure, one must specify the temperature and magnetic field present in the region between the plates and in the
region outside the plates,
since the Casimir pressure depends on the conditions of both regions. Terms in the free energy that are proportional to $a$ are uniform
energy density terms, and will not contribute to the pressure if the medium outside the plates is at the same temperature
and with the same magnetic field as the medium between the plates. I will assume that the medium inside and outside the plates is at the same temperature, and will investigate the cases when the magnetic field is only present between the plates, and when it is present between and outside the plates.

For small $a$, I use (\ref{F0})  into (\ref{P1}) and find
$$P_1={7\pi^2\over 1,920 a^4}-{e^2B^2\over 48\pi^{2}}\left(1+\gamma_E + \ln{2a m_\phi \over\pi}\right)+
{ eB \over 4a^2} 
{ e^{-\pi \beta/2a}\over \sinh ({eBa\beta\over \pi})},
$$
when there is magnetic field only between the plates,
and
$$P_2={7\pi^2\over 1,920 a^4}-{e^2B^2\over 48\pi^{2}}\left(1+ \ln{2a m_\phi \over\pi}\right)+
{ eB \over 4a^2} 
{ e^{-\pi \beta/2a}\over \sinh ({eBa\beta\over \pi})},
$$
for a magnetic field present inside and outside the plates. In both cases I neglected some smaller terms
that do not contribute significantly to the pressure. The pressure is 
repulsive in both cases, as expected. Notice that the presence of a magnetic field between the plates, but not outside, 
weakens the repulsive pressure, since $\Delta P = P_1-P_2$ is
$$\Delta P=-{\gamma_E\over 48\pi^{2}}e^2B^2.$$

In the limit of high temperature and $\sqrt{eB}\gg a^{-1}$, I use (\ref{zBa020a}) and (\ref{P1}) to find
$$P_1=
{(eB)^{3/2}\over 2\pi\beta}(\sqrt{2}-1)\zeta_R(-{\textstyle\frac{1}{2}})
-{e^2B^2\over 12\pi}\left(\gamma_E+\ln{ m_\phi \beta\over4 \pi}\right)
+{(eB)^{3/2}\over \pi\beta }e^{-2a\sqrt{eB}}
+2{eBe^{-4\pi a/\beta}\over \beta^2 \sinh ({eBa\beta\over 2\pi})},
$$
for the case when the magnetic field is present only between the plates, and 
$$P_2=
{(eB)^{3/2}\over \pi\beta }e^{-2a\sqrt{eB}}
+2{eBe^{-4\pi a/\beta}\over \beta^2 \sinh ({eBa\beta\over 2\pi})},
$$
when the magnetic field is present between the plates and outside. I use (\ref{zBa021a}), (\ref{P1}) and obtain
the high temperature limit in the case of $a^{-1}\gg\sqrt{eB} $, 
$$P_1=
{(eB)^{3/2}\over 2\pi\beta}(\sqrt{2}-1)\zeta_R(-{\textstyle\frac{1}{2}})
-{e^2B^2\over 12\pi}\left({a\over 2\beta}+\gamma_E+\ln{ m_\phi \beta\over4 \pi}\right)+
{3\zeta_R(3)\over 16\pi \beta a^3}+
\left({4\pi\over a\beta^3}-{e^2B^2a\over 6\pi\beta}\right){ e^{-4\pi a/\beta}},
$$
for a magnetic field present only between the plates, and
$$P_2=
-{e^2B^2a\over 24\pi\beta}+
{3\zeta_R(3)\over 16\pi \beta a^3}+
\left({4\pi\over a\beta^3}-{e^2B^2a\over 6\pi\beta}\right){ e^{-4\pi a/\beta}},
$$
when a magnetic field is present between the plates and outside. In both cases 
I neglected some smaller terms that do not contribute significantly to the pressure. 
Notice that the dominant term of the high
temperature free energy (\ref{zBa020a}) and (\ref{zBa021a}) is the Stefan-Boltzman term, ${\pi^2\over 90\beta^4}$,
but it does not contribute to the pressure because it is a uniform energy density term dependent on the temperature only. 
In the high temperature limit
\begin{equation}
\Delta P={(eB)^{3/2}\over 2\pi\beta}(\sqrt{2}-1)\zeta_R(-{\textstyle\frac{1}{2}})
-{e^2B^2\over 12\pi}\left(\gamma_E+\ln{ m_\phi \beta\over4 \pi}\right),
\label{p1p21}
\end{equation}
for both $\sqrt{eB}\gg a^{-1}$and $a^{-1}\gg\sqrt{eB} $. Since $\zeta_R(-{\textstyle\frac{1}{2}})$ is negative, the presence of a magnetic field 
between the plates, but not outside, weakens the repulsive pressure also in the high temperature limit, and can even reverse the pressure to an attractive one. 

Finally, in the strong magnetic field limit I use (\ref{zBB7}), (\ref{P1}) and find
$$
P_1= {(eB)^2\over 96 \pi^2}\left(\ln{eB\over 3  m_\phi^2}-{1\over 2}\right)+
{(eB)^{5/4}\over\sqrt{2}(\pi\beta)^{3/2}}e^{-\beta\sqrt{eB}}+
{(3eB)^{5/4}\over\sqrt{2}(\pi\beta)^{3/2}}e^{-\beta\sqrt{3eB}}+
{(eB)^{7/4}\over 2\pi^{3/2}\sqrt{a}}
e^{-2a\sqrt{eB}},
$$
$$
P_2= {(eB)^{7/4}\over 2\pi^{3/2}\sqrt{a}}
e^{-2a\sqrt{eB}},
$$
and
$$\Delta P={(eB)^2\over 96 \pi^2}\left(\ln{eB\over 3  m_\phi^2}-{1\over 2}\right)+
{(eB)^{5/4}\over\sqrt{2}(\pi\beta)^{3/2}}e^{-\beta\sqrt{eB}}+{(3eB)^{5/4}\over\sqrt{2}(\pi\beta)^{3/2}}e^{-\beta\sqrt{3eB}},
$$
where the second and third term, exponentially suppressed for $\sqrt{eB}\beta\gg 1$, are in 
most cases much smaller than the first term. 
The dominant first term, and thus 
$\Delta P$, can be positive or negative depending on the value of the strong magnetic field $B$. Therefore the 
presence of a magnetic field between the plates but not outside, will increase the repulsive pressure in the case
of a very strong magnetic field, and will decrease it in the case of a moderately strong field.
\section{Discussion and conclusions}
\label{7}
In this paper I used the zeta function technique to study magnetic effects on the repulsive Casimir effect at finite
temperature. I investigated a massless and charged scalar field satisfying mixed boundary conditions on two 
parallel plates, in thermal equilibrium with a heat reservoir and in the presence of a uniform magnetic field perpendicular
to the plates. I obtained  simple analytic expressions for the Helmoltz free energy  in the case of small plate distance (\ref
{F0}), high temperature (\ref{zBa020a}) and (\ref{zBa021a}), and strong magnetic field (\ref{zBB7}).

Using the method described in Refs. \cite{Erdas:2013jga,Erdas:2013dha} and based upon the fact that, for a well behaved
function $G(s)$, the derivative of 
$G(s)/\Gamma(s)$ at $s=0$ is simply $G(0)$, I obtained exact expressions of the 
free energy suitable for accurate numerical evaluation in each of the three limits.  I compared the values of the free energy obtained from
my simple analytic expressions to the exact numerical values. In the small plate distance case I find that, for $aT\le {1\over 2}$ 
and $a\sqrt{eB}\le {1\over 2}$,
Eq. (\ref{F0}) is within $3.3$ percent of the exact value of the free energy, while for $aT\le {1\over 5}$ 
and $a\sqrt{eB}\le {1\over 5}$,
Eq. (\ref{F0}) is within $0.001$ percent of the exact free energy, showing that the discrepancy with the exact value falls
very rapidly as $aT$ and $a\sqrt{eB}$ decrease. The analytic expressions (\ref{zBa020a}) and 
(\ref{zBa021a}) for the
high temperature limit of the free energy are considerably more accurate, since for $Ta\ge 2$ and $T/\sqrt{eB}\ge 2$ their discrepancy with the exact value of the free energy is less than
0.1 percent, and falls very rapidly as $Ta$ and $T/\sqrt{eB}$ grow. Finally, in the strong magnetic field limit, the free energy of Eq. (\ref{zBB7}) is within 0.2 percent of the exact value for $\sqrt{eB}a\ge 2$ and $\sqrt{eB}/T\ge 2$, with an accuracy
similar to that of Eqs. (\ref{zBa020a}) and (\ref{zBa021a}).

The simple analytic expressions of the Casimir pressure in the three limits, obtained in Sec. \ref{6}, are as accurate as those of the free energy.
It is discovered that, in the small plate distance and high temperature limit, the repulsive Casimir pressure is always weaker when a 
magnetic field is present between the plates but not outside, than it is when a magnetic field is present between the plates and outside.
In the strong magnetic field limit this effect happens only for a moderately strong field, while the repulsion increases when a very strong 
magnetic field is examined.

\end{document}